# Spiral Capacitor Calculation Using FEniCS

Slava Andrejev

February 12, 2019

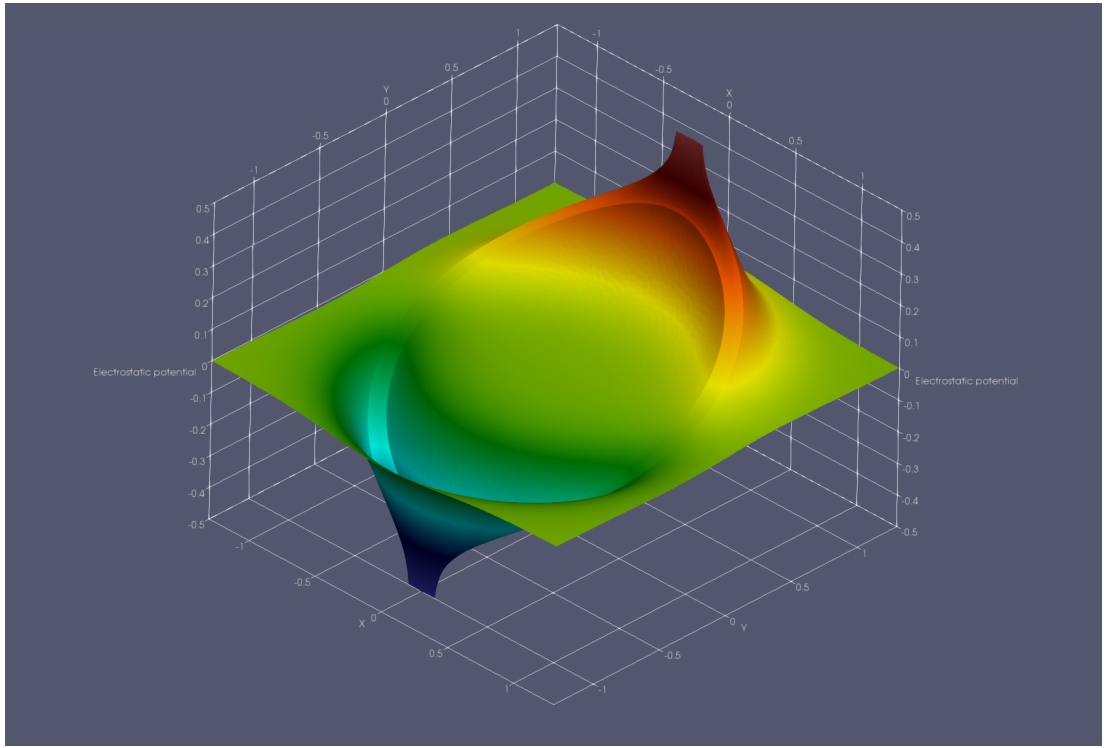

The paper shows how to optimize a water level sensor consisting of a cylinder with spiraling metal stripes on the side, using a powerful Python library FEniCS. It is shown how to reduce a 3D Laplace equation to a 2D, using a spiraling coordinate system; how to specify the correct boundary conditions for an open region; how to convert the partial differential equation to a variational form for FEniCS; and how to calculate the capacitance. Then the FEniCS code is shown that solves the Laplace equation and calculates the capacitance. The further numeric experiments show that there is an optimal combination of the spiral frequency and the width of the stripes that maximizes the sensitivity of the sensor. The Python code is given to calculate the optimum.

## What is a spiral capacitor and why one might need it?

Don't google "spiral capacitor" I came up with this term to call what I have done for a DIY project. A spiral capacitor is a cylinder with two stripes of metal foil spiraling across each other along the cylinder length. That's it. But, well, it doesn't take the whole section. Ok, the real reason behind this section is to explain why I have written this article and maybe get you a bit excited about what you are going to learn from it.

I was contemplating on how to make a water level sensor for a DIY project. One of the ideas was to place two stripes of foil on the sides of a cylinder that holds water. When the level of water changes, it changes the electric capacitance between the stripes, which then can be measured. Then I thought: if I wind the stripes around the cylinder in a spiral, I probably will increase the capacitance and probably the sensitivity? However, I didn't have answers to a lot of pragmatic questions. What will be the capacitance? Will it change significantly with the water level? How it will change with the spiral angle? What is the best strategy: make narrow stripes and wind them tightly or make wide stripes and wind them with a moderate spiral? Indeed, one may think that if we tightly wind two narrow stripes, each loop will interact more with loops above and below than across the cylinder, thus making it insensitive to the water inside. Therefore, maybe there is some optimal angle of the spiral?

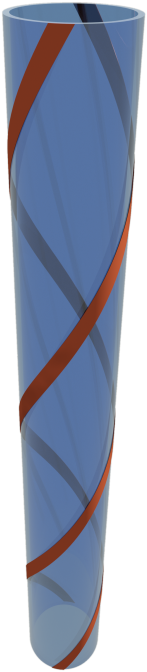

**Figure 1:** Spiral capacitor model.

All these questions would be very time and material consuming to answer by experimenting with the real cylinders and spirals. To solve the problem numerically one has to solve a partial differential equation. Partial differential equations (PDE) is a complicated mathematical topic. The complicity is regrettable, considering PDEs describe a plethora of phenomena around us that we encounter every second, whether we drink coffee or play a drum. Fortunately, nowadays we have in our possession a lot of tools that simplify hundredfold the task of solving PDE and help getting answers to practical questions that you can use even at your home, for DIY or just for fun. For example, how much more dollars you will spend on heating if you make another window in the north wall of your house?

I was so fascinated when I found a great tool that made the work very easy. This is a Python library called FEniCS [1, 2]. It was so great that I decided to write a paper about my endeavor with calculating this project. This library solves PDE using the finite element method (FEM). The learning curve is still steep, but once you get it, it's no brainer and is a lot of fun. There are a lot of tutorials how to use FEniCS, but most of them are not related to practice. Indeed, how you are going to use in practice a solution of a Poisson equation $\nabla^2 u = 6$? This is the problem from which the FEniCS tutorial starts. So, I picked a problem that I in fact encountered in practice and would like to share how to solve it.

In the first several sections there will be a lot of math. It describes how to transform the problem so that it is easier to solve numerically, and plus it describes what we need to calculate to get our practical answers. The math though is not difficult, the first year of calculus will do to understand. Nowadays you can also find everything in Wikipedia and YouTube. These sections also show important techniques you can use yourself.

There are a few notations and terminology from differential geometry. Vectors are of two different kinds: contravariant and covariant. Indices of contravariant vectors are up, for example $v^i$, correspondingly covariant vector components have their indices down: $v_i$. The same rule is applied to any tensor: contravariant indices are up, covariant down. Einstein summation convention is assumed: if an index is repeated in one term up and down, it is equivalent of summation over this index. Examples: $x_i y^i$ is a scalar product, $\nabla_i v^i$ is a divergence in the Cartesian coordinates, and $\delta^i_j x^j = x^i$.

Below is a brief summary of the notations used in the following sections.

| Symbol | Meaning |
|---|---|
| $\nabla$ | Nabla symbol, a vector of partial derivatives: $\frac{\partial}{\partial x^i}$. |
| $\nabla^2$ | Laplace operator: $\sum_i \frac{\partial^2}{\partial {x^i}^2}$. |
| $\nabla_i$ | Partial derivative $\frac{\partial}{\partial x^i}$. |
| $\partial_x$ | Short for $\frac{\partial}{\partial x}$. |
| $g_{ij}$ | Metric tensor. It describes how to calculate a scalar product and a vector length in a curvilinear coordinate system: $x \cdot y = x^i y^j g_{ij}$, $|x|^2 = x^i x^j g_{ij}$. In the Cartesian coordinates $g_{ij}$ is obviously an identity matrix, i.e. only diagonal elements are not zero, and all diagonal elements are 1. |
| $g^{ij}$ | Contravariant metric tensor. It's equal to $g_{ij}^{-1}$. |
| $|g|$ | Metric tensor determinant. |
| $\delta^{ij}$ | Kronecker delta. Only diagonal elements are not zero, and all diagonal elements are 1. |
| $\varepsilon_i^j$ | Levi-Civita symbol. |
| $\Omega$ | A region in space. |
| $\partial\Omega$ | The boundary of $\Omega$. |
| $u$ | 3D electrostatic potential. |
| $u_{2D}$ | 2D slice of $u$ across the axis of the cylinder. |
| $E$ | Electric field. |
| $D$ | Electric displacement field. |
| $\epsilon$ | Dielectric permittivity. |

# Electrostatic potential

An electrostatic potential is the fundamental scalar quantity that starts solution of any electrostatic task. If we calculate it, we can find everything else: the electric field, energy, capacitance. We will denote it with a letter $u$. In the region of space without electric charges it is described by a Laplace equation:

$$\nabla^2 u = 0.$$

Let's define a spatial frequency $\omega$ of the winding metal stripes: the angle our spiral turns around the cylinder axis per unit of length of the axis. We can reduce 3D problem to 2D if we introduce a coordinate system that rotates with the spiral. Thus, the Laplace equation will not depend on where along the axis of the cylinder we solve it, and we can solve a 2D equation instead of 3D. Let's choose our conventional Cartesian $x$, $y$, $z$ coordinate system so that $Z$ axis is the axis of the cylinder. For the winding coordinate system we choose a cylindrical coordinate system in which $\varphi$ shifts to angle $\omega z$, thus rotating the 2D polar coordinates $r$, $\varphi$ with the spiral. We could choose a winding coordinate system that is Cartesian in $XY$ plane and also rotates around $Z$ to the angle $\omega z$, but we will later see that the polar coordinates will allow us to easily recognize a divergence of a specially constructed field in the Laplacian. The equation that transforms rotating cylindrical coordinates to normal Cartesian looks like this

$$\begin{cases} x = r\cos(\varphi + \omega z) \\ y = r\sin(\varphi + \omega z) \\ z = z \end{cases}. \tag{1}$$

This corresponds to a spiral that winds counter clockwise if we look down the negative direction of $Z$ axis. To use formulae of the differential geometry we need coordinate components with indices. For example, let $x, y, z$ be $\eta^1, \eta^2, \eta^3$, and $r, \varphi, z$ be $\zeta^1, \zeta^2, \zeta^3$.

Let's find the Laplacian in the curvilinear coordinates (1). In an arbitrary curvilinear coordinates the

Laplacian can be expressed using the metric tensor components:

$$\nabla^2 u = \frac{1}{\sqrt{|g|}} \frac{\partial}{\partial \zeta^i}\left(\sqrt{|g|} g^{ij} \frac{\partial u}{\partial \zeta^j}\right).$$

This equation assumes that $g$ is so called *induced* metric tensor, i.e. it has an explicit form written via coordinates derivatives. Considering that for a Cartesian coordinate system $g_{ij} = \delta_{ij}$, the induced metric tensor for a curvilinear coordinate system can be written as

$$g_{km} = \delta_{ij} \frac{\partial \eta^i}{\partial \zeta^k} \frac{\partial \eta^j}{\partial \zeta^m}. \tag{2}$$

The induced metric tensor (2) can lead to confusion. For example, if you try to use it for polar coordinates, you will find out that it assumes the $\varphi$ component of a vector field is measured in radians, while in the canonical polar coordinates that we get accustomed to, it is the length units, same as the radius component.

Substituting (1) to (2) will produce

$$g_{ij} = \begin{pmatrix} 1 & 0 & 0 \\ 0 & \zeta^{1^2} & \omega \zeta^{1^2} \\ 0 & \omega \zeta^{1^2} & 1 + \omega^2 \zeta^{1^2} \end{pmatrix},$$

with the inverse

$$g^{ij} = g_{ij}^{-1} = \begin{pmatrix} 1 & 0 & 0 \\ 0 & \omega^2 + \dfrac{1}{\zeta^{1^2}} & -\omega \\ 0 & -\omega & 1 \end{pmatrix},$$

and the square root of the determinant

$$\sqrt{|g|} = \zeta^1.$$

Therefore, the Laplacian of an arbitrary function $f$ in this coordinate system will be

$$\nabla^2 f = \frac{1}{\zeta^1} \frac{\partial}{\partial \zeta^1}\left(\zeta^1 \frac{\partial f}{\partial \zeta^1}\right) + \frac{1 + \omega^2 \zeta^{1^2}}{\zeta^{1^2}} \frac{\partial^2 f}{\partial \zeta^{2^2}} - 2\omega \frac{\partial^2 f}{\partial \zeta^2 \partial \zeta^3} + \frac{\partial^2 f}{\partial \zeta^{3^2}}.$$

We are going to calculate the electrostatic potential $u$. In this coordinate system it will not depend on $\zeta^3$, therefore, we can get rid of $\zeta^3$ derivatives in the Laplace equation:

$$\frac{1}{\zeta^1} \frac{\partial}{\partial \zeta^1}\left(\zeta^1 \frac{\partial u_{2D}}{\partial \zeta^1}\right) + \frac{1 + \omega^2 \zeta^{1^2}}{\zeta^{1^2}} \frac{\partial^2 u_{2D}}{\partial \zeta^{2^2}} = 0.$$

Index $2D$ means it's a two-dimensional slice of the full three-dimensional potential. At this point we can safely rewrite it using regular polar coordinates:

$$\frac{1}{r} \frac{\partial}{\partial r}\left(r \frac{\partial u_{2D}}{\partial r}\right) + \frac{1 + \omega^2 r^2}{r^2} \frac{\partial^2 u_{2D}}{\partial \varphi^2} = 0. \tag{3}$$

These polar coordinates don't suffer from the choice of equation (2) for a metric tensor, because there are no vector fields in this equation. Let's introduce a notation $\nabla^2_\omega f$ of a Laplacian with an index $\omega$:

$$\nabla^2_\omega f = \frac{1}{r} \frac{\partial}{\partial r}\left(r \frac{\partial f}{\partial r}\right) + \frac{1 + \omega^2 r^2}{r^2} \frac{\partial^2 f}{\partial \varphi^2}$$

to distinguish it from a regular Laplacian in the polar coordinates

$$\nabla^2 f = \frac{1}{r}\frac{\partial}{\partial r}\left(r\frac{\partial f}{\partial r}\right) + \frac{1}{r^2}\frac{\partial^2 f}{\partial \varphi^2}.$$

Obviously,

$$\nabla_\omega^2 f = \nabla^2 f + \omega^2 \frac{\partial^2 f}{\partial \varphi^2}.$$

Note that now equation (3) or its equivalent $\nabla_\omega^2 u_{2D} = 0$ can be written in the following interesting form:

$$\nabla^2 u_{2D} = -\omega^2 \frac{\partial^2 u_{2D}}{\partial \varphi^2}, \tag{4}$$

which should look familiar: it is similar to a Poisson equation, at least its left-hand side. The right-hand part should correspond to the charge density. Therefore, a 2D slice of our electrostatic potential behaves as if there is a dynamic charge distribution in the space. We will later see that this will lead to a very important and counterintuitive at first result.

For the finite element method, we should rewrite equation (3) as a divergence of some vector field in the polar coordinates. This is where the choice we mentioned before of the winding cylindrical coordinates instead of a Cartesian system gets handy, because it's easy to see the polar coordinate divergence formula in (3):

$$\nabla_\omega^2 u_{2D} = \nabla \begin{pmatrix} \dfrac{\partial u_{2D}}{\partial r} \\ \dfrac{1+\omega^2 r^2}{r}\dfrac{\partial u_{2D}}{\partial \varphi} \end{pmatrix} = 0. \tag{5}$$

This vector field has nothing to do with the electrical field, it's just a mathematical trick to rewrite the equation. Let's denote this vector field as $\nabla_\omega$:

$$\nabla_\omega u_{2D} = \begin{pmatrix} \dfrac{\partial u_{2D}}{\partial r} \\ \dfrac{1+\omega^2 r^2}{r}\dfrac{\partial u_{2D}}{\partial \varphi} \end{pmatrix}.$$

Equation (5) uses a traditional polar coordinate system, i.e. it is orthonormal, and both $r$ and $\varphi$ components of a vector field have units of length. If we used the induced metric tensor (2), the $\varphi$ component would be equal to $(1+\omega^2 r^2)r^{-2}\partial_\varphi u_{2D}$. Now we can convert (5) back to the Cartesian coordinates:

$$\nabla \begin{pmatrix} \left(1+\omega^2 {x^2}^2\right)\dfrac{\partial u_{2D}}{\partial x^1} - \omega^2 x^1 x^2 \dfrac{\partial u_{2D}}{\partial x^2} \\ -\omega^2 x^1 x^2 \dfrac{\partial u_{2D}}{\partial x^1} + \left(1+\omega^2 {x^1}^2\right)\dfrac{\partial u_{2D}}{\partial x^2} \end{pmatrix} = 0. \tag{6}$$

Note that these 2D Cartesian coordinates $x^1$ and $x^2$ are not our $x$, $y$ from the stationary main coordinate system. $x^1$ and $x^2$ are winding with the spiral, i.e. $u(x^1, x^2)$ does not depend on $z$ coordinate. It's obvious from (6) that our vector field is an ordinary gradient plus same gradient multiplied by a matrix proportional to $\omega^2$. This matrix is an outer product of a vector $[x^2, -x^1] = \varepsilon_k^i x^k$. Therefore, we can further simplify (6):

$$\nabla_i \left(\delta^{ij} + \omega^2 \varepsilon_k^i \varepsilon_l^j x^k x^l\right) \nabla_j u_{2D} = 0. \tag{7}$$

# Capacitance

To calculate the capacitance, we need to know the energy stored in a capacitor. The energy can be calculated by integrating the electric field energy density over the capacitor volume. The electric field energy density is

$$\frac{\mathbf{D} \cdot \mathbf{E}}{8\pi},$$

where

$$\mathbf{E} = -\nabla u,$$

and $\mathbf{D}$ is the electric displacement field, which in our case is defined as

$$\mathbf{D} = \epsilon \mathbf{E},$$

where $\epsilon$ is dielectric permittivity. Putting all together, the stored energy is

$$E = \frac{\epsilon}{8\pi} \iiint_\Omega |\nabla u|^2 \, dx dy dz. \tag{8}$$

We need to be careful with calculation of $\nabla u$. In the equation above it means a 3D gradient. However, $u_{2D}$ that we find by solving (6) is a 2D function. We get a 3D potential by rotating $u_{2D}$ for every coordinate $z$ by angle $\omega z$:

$$u(x, y, z) = u_{2D}(x \cos \omega z + y \sin \omega z, -x \sin \omega z + y \cos \omega z).$$

Therefore, after performing all partial derivative calculations:

$$|\nabla u|^2 \bigg|_{z=0} = |\nabla u_{2D}|^2 + \omega^2 \left( x^2 \frac{\partial u_{2D}}{\partial x^1} - x^1 \frac{\partial u_{2D}}{\partial x^2} \right)^2.$$

Or, using the index notation

$$|\nabla u|^2 \bigg|_{z=0} = |\nabla u_{2D}|^2 + \omega^2 \left( \varepsilon^i_j x^j \partial_i u_{2D} \right)^2. \tag{9}$$

If we change variables in (8) from $x$, $y$, $z$ to $x^1$, $x^2$, $z$ (remember $x^1$ and $x^2$ wind with the spiral), $|\nabla u|^2$ will not depend on $z$ coordinate, therefore (8) can be reduced to a two-dimensional integral for stored energy per unit of length. The Jacobian of the transition from $x$, $y$, $z$ to $x^1$, $x^2$, $z$ is equal 1, so we can just substitute (9) to (8) and remove one integral sign:

$$E_{2D} = \frac{\epsilon}{8\pi} \iint_{\Omega_{2D}} \left( |\nabla u_{2D}|^2 + \omega^2 \left( \varepsilon^i_j x^j \nabla_i u_{2D} \right)^2 \right) d\mathbf{x}. \tag{10}$$

Finally, the capacitance is

$$C = \frac{2 E_{2D}}{V^2}, \tag{11}$$

where $V$ is the voltage applied to the capacitor.

# Boundary conditions for infinity

The final preparation step before we can use a numerical method to solve (3) is to find the correct boundary conditions. Our problem is open, meaning that we don't have a compact area in which we are solving (3) with a prescribed value of potential in each point on the boundary. Instead, we are solving a second order PDE in the whole space with the values of potential known only on the plates of our capacitor. Of course, we cannot numerically solve (3) in the whole space. Instead, we can try to take a large enough circle and specify an approximate boundary condition on its edge. In order to do this, we need to know how the solution of (3) behaves far away from the cylinder.

At the distances big enough compared to the radius of the cylinder the potential will be similar to potential of a two oppositely charged wires tightly twisted with each other. I.e. we have to find a dipole potential for (3).

Let's take equation (3) and write a solution in an analytic form. Using the method of separation of variables with a proposed solution $u_{2D}(r,\varphi) = R(r)T(\varphi)$ in (3), produces

$$\frac{d^2R(r)}{dr^2}\frac{1}{R(r)}\frac{r^2}{1+\omega^2 r^2} + \frac{dR(r)}{dr}\frac{1}{R(r)}\frac{r}{1+\omega^2 r^2} = -\frac{d^2T(\varphi)}{d\varphi^2}\frac{1}{T(\varphi)},$$

which is fully separated in $r$ and $\varphi$. Because the left-hand side depends only on $r$, and the right-hand side depends only on $\varphi$, they both are equal to some constant. $T(\varphi)$ should be a periodic function, which means the right-hand side is an oscillator equation:

$$-\frac{d^2T(\varphi)}{d\varphi^2}\frac{1}{T(\varphi)} = n^2,$$

with solutions

$$T_n(\varphi) = \alpha_n \cos n\varphi + \beta_n \sin n\varphi.$$

For the left-hand side we get the following equation:

$$r^2\frac{d^2R(r)}{dr^2} + r\frac{dR(r)}{dr} - n^2\left(1+\omega^2 r^2\right)R(r) = 0.$$

Changing variable $z = n\omega r$ will transform it to

$$z^2\frac{d^2R}{dz^2} + z\frac{dR}{dz} - \left(n^2 + z^2\right)R = 0.$$

Solutions of this equation are modified Bessel functions of the first and second kind: $I_n$ and $K_n$. $I_n(z)$ are infinite at $z = \infty$, therefore $R$ can be only a linear combination of $K_n$. A generic solution is

$$u_{2D}(r,\varphi) = \frac{\alpha_0}{2} + \sum_{n=1}^{\infty} K_n(n\omega r)\left(\alpha_n \cos n\varphi + \beta_n \sin n\varphi\right). \tag{12}$$

Let's assume our plates are placed symmetrically on both sides of $X$ axis, then $u_{2D}$ will be an odd function with respect to $\varphi$, and therefore $\alpha_i = 0 \; \forall i$ in (12):

$$u_{2D}(r,\varphi) = \sum_{n=1}^{\infty} K_n(n\omega r)\beta_n \sin n\varphi.$$

The very first term of this series is our dipole potential:

$$W(r, \varphi) = \beta_1 K_1(\omega r) \sin \varphi. \tag{13}$$

It's an amazing result, because $K_1(\omega r)$ has an exponential decay law at the large distances. Usually electrostatic tasks give us an inverse power decay, however, here a simple twist of a pair of stripes has so excellent self-screening property that it leads to a much faster, exponential decay. It's a property usually observed only if there are free charges that screen the potential. This is the time to recall that equation (4) reminds a Poisson equation with free charges.

To be sure that the result is correct, we can get a similar outcome by integrating an infinite wire consisting of $3D$ dipoles directed perpendicular to the wire and rotating around the wire with spatial frequency $\omega$. A dipole direction will be a $3D$ vector with coordinates $\sin \omega z, \cos \omega z$, and $0$. A vector from the point of interest $(r, \omega)$ to the dipole is $(r \cos \varphi, r \sin \varphi, -z)$. Therefore, the dipole potential at the point of interest, omitting a multiplicative constant, will be

$$\frac{r \sin(\omega z + \varphi)}{(z^2 + r^2)^{3/2}}.$$

If we integrate the above over $z$, we will get the same result as (13):

$$\int_{-\infty}^{\infty} \frac{r \sin(\omega z + \varphi)}{(z^2 + r^2)^{3/2}} \, dz = 2\omega K_1(\omega r) \sin \varphi.$$

We are ready to get the boundary conditions. Considering

$$\frac{\partial K_1(\omega r)}{\partial r} = -\frac{\omega}{2}\left(K_0(\omega r) + K_2(\omega r)\right),$$

we can write down a Rubin boundary condition for a circle with a radius $R$ significantly larger than the radius of our cylinder:

$$u_{2D} + \frac{\partial u_{2D}}{\partial r} \frac{2 K_1(\omega R)}{\omega (K_0(\omega R) + K_2(\omega R))} = 0. \tag{14}$$

Let's check that for small $\omega$ we will get a boundary condition for a $2D$ dipole. For small $\omega$, i.e. when $\omega R \ll 1$,

$$\frac{2 K_1(\omega R)}{\omega (K_0(\omega R) + K_2(\omega R))} = R + \mathcal{O}\left((\omega R)^2\right).$$

If we substitute this to (14), we will get a boundary condition for an ordinary $2D$ dipole:

$$u_{2D} + R \frac{\partial u_{2D}}{\partial r} = 0.$$

## Numeric analysis using FEniCS

The following analysis is not concerned with the quantities expressed as meters or farads. We are interested only in the qualitative results, i.e. questions like how the capacitance changes with the winding angle, what is the optimal angle, how to achieve the maximum sensitivity? Therefore, the radius of the cylinder is just one of some units of length. It can be one of millimeters, feet, miles, or $\ln^{\gamma} \pi$ yards. Therefore, we won't calculate the capacitance in farads, but just express as the stored energy. The energy is also expressed in some arbitrary units, they will depend on what the length units are, but this is beyond the scope of this paper.

There are a lot of excellent resources online to get you started with FEniCS. FEniCS installation instructions [3] describe how to quick start on several platforms. An excellent tutorial book is available for free online [4].

## Variational formulation

In order to use FEniCS we have to write the equation we are trying to solve in a so-called variational form:

$$a(u, v) = L(v) \quad \forall v.$$

The unknown function is $u$, $v$ is called a *trial function*, a placeholder to state that the above equation holds for any $v$. $a(u, v)$ is a *bilinear form*, $L(v)$ is a *linear form*. To get a variational form we have to consider a more general equation than (5)

$$\nabla_\omega^2 u_{2D} = \nabla \begin{pmatrix} \dfrac{\partial u_{2D}}{\partial r} \\ \dfrac{1 + \omega^2 r^2}{r} \dfrac{\partial u_{2D}}{\partial \varphi} \end{pmatrix} = -f, \tag{15}$$

where $f$ is some known scalar function of coordinates. Later we will make it equal to zero, but it is essential for FEniCS to figure out what it has to solve. Let's use the Gauss theorem in the following form:

$$\int_\Omega \frac{1}{\sqrt{|g|}} \partial_i \left(\sqrt{|g|} A^i\right) dx = \oint_{\partial\Omega} n^j A^i g_{ij} \, ds. \tag{16}$$

The left integral is an integral of the divergence of a vector field **A** expressed in curvilinear coordinates, the right integral is an integral of **A** over the boundary of $\Omega$. Let's assume **A** is the following expression:

$$\mathbf{A} = v \begin{pmatrix} \dfrac{\partial u_{2D}}{\partial r} \\ \dfrac{1 + \omega^2 r^2}{r^2} \dfrac{\partial u_{2D}}{\partial \varphi} \end{pmatrix}, \tag{17}$$

where $v$ is a scalar function of coordinates. Note $r^{-2}$, not $r^{-1}$ in the $\varphi$ component. It's different from (15), because (16) uses an induced metric tensor. See notes for equation (5). Substituting (17) to (16) will produce

$$\int_\Omega \left(v \nabla_\omega^2 u_{2D} + \partial_r v \partial_r u_{2D} + \partial_\varphi v \partial_\varphi u_{2D} \frac{1 + \omega^2 r^2}{r^2}\right) dx = \oint_{\partial\Omega} v \left(n_r \partial_r u_{2D} + n_\varphi \frac{1 + \omega^2 r^2}{r} \partial_\varphi u_{2D}\right) ds,$$

where $n_r$ and $n_\varphi$ are components of the $\partial\Omega$'s normal. It's not difficult to see a scalar product of $\nabla v$ and $\nabla_\omega u$ in the left part and a scalar product $\mathbf{n} \cdot \nabla_\omega u$ in the right:

$$\int_\Omega \left(v \nabla_\omega^2 u_{2D} + \nabla v \cdot \nabla_\omega u_{2D}\right) dx = \oint_{\partial\Omega} v \mathbf{n} \cdot \nabla_\omega u_{2D} \, ds.$$

Since $\nabla_\omega^2 u_{2D} = -f$

$$\int_\Omega \nabla v \cdot \nabla_\omega u_{2D} \, dx = \int_\Omega v f \, dx + \oint_{\partial\Omega} v \mathbf{n} \cdot \nabla_\omega u_{2D} \, ds.$$

The boundary $\partial\Omega$ consists of three parts: two arcs on the sides of a circle, corresponding to the metal on the sides of the cylinder, and third one is the outside edge of our whole region where we are looking for a solution. If we take a large circle of radius $R$ as $\partial\Omega$, then the boundary condition (14) will hold. Let's denote the edge of this circle as $\partial\Gamma_R$. On the metal plates the surface integral will vanish, because for a conductor we have a Dirichlet boundary condition, and for a Dirichlet boundary condition $v = 0$. Therefore,

$$\int_\Omega \nabla v \cdot \nabla_\omega u_{2D}\, fx = \int_\Omega vf\, dx + \oint_{\partial\Gamma_R} v\, \mathbf{n} \cdot \nabla_\omega u_{2D}\, ds.$$

Substituting $\mathbf{n} \cdot \nabla_\omega u$ from (14) will give us

$$\int_\Omega \nabla v \cdot \nabla_\omega u\, dx + \oint_{\partial\Gamma_R} vu_{2D} \frac{\omega\left(K_0(\omega R) + K_2(\omega R)\right)}{2K_1(\omega R)}\, ds = \int_\Omega vf\, dx.$$

This is our variational form. Therefore,

$$a(u, v) = \int_\Omega \nabla v \cdot \nabla_\omega u\, dx + \oint_{\partial\Gamma_R} vu_{2D} \frac{\omega\left(K_0(\omega R) + K_2(\omega R)\right)}{2K_1(\omega R)}\, ds, \qquad (18)$$

and

$$L(v) = \int_\Omega vf\, dx. \qquad (19)$$

Let's briefly reiterate what we have done. We started from the Gauss theorem for our specially designed vector field on the right-hand side of (5) multiplied by a trial function. Then we substituted there: (i) equation (15) we have to solve; (ii) the Rubin boundary condition (14). The result is the variational form of our problem expressed via functions (18) and (19).

### Creating a mesh

To solve a PDE we have to discretize our 2D space. A usual way to do it for FEM is triangulation. There is a catch though. The trick is that the triangle edges have to be aligned with the boundaries of our problem. Also, it would be nice to have denser triangulation in areas where the potential has more details, i.e. near the metal plates. This is not a trivial task. Fortunately, there is a script language Gmsh [5, 6] that can generate complicated meshes. Below is a script that generates a mesh for our problem.

```
 1   centerDensity = 0.1;
 2   cylDensity = 0.01;
 3   outDensity = 0.25;
 4   nearCylDensity = 0.02;
 5   cylSize = 1;
 6   boundarySize = 5;
 7   dSize = 0.1;
 8   wallSize = 0.1;
 9   centerSize = 0.2;
10
11   Point(1) = {0, 0, 0, centerDensity};
12
13   Point(2) = { centerSize,            0, 0, centerDensity};
14   Point(3) = {           0,  centerSize, 0, centerDensity};
15   Point(4) = {-centerSize,            0, 0, centerDensity};
16   Point(5) = {           0, -centerSize, 0, centerDensity};
17
18   Point(6) = { cylSize,           0, 0, cylDensity};
19   Point(7) = {       0,    cylSize, 0, cylDensity};
20   Point(8) = {-cylSize,           0, 0, cylDensity};
21   Point(9) = {       0,   -cylSize, 0, cylDensity};
22
23   Point(10) = { boundarySize,              0, 0, outDensity};
24   Point(11) = {            0,   boundarySize, 0, outDensity};
25   Point(12) = {-boundarySize,              0, 0, outDensity};
26   Point(13) = {            0,  -boundarySize, 0, outDensity};
```

```
27
28   Point(14) = { cylSize + dSize,               0, 0, nearCylDensity};
29   Point(15) = {              0,  cylSize + dSize, 0, nearCylDensity};
30   Point(16) = {-cylSize - dSize,               0, 0, nearCylDensity};
31   Point(17) = {              0, -cylSize - dSize, 0, nearCylDensity};
32
33   Point(18) = { cylSize - wallSize,              0, 0, nearCylDensity};
34   Point(19) = {              0,  cylSize - wallSize, 0, nearCylDensity};
35   Point(20) = {-cylSize + wallSize,              0, 0, nearCylDensity};
36   Point(21) = {              0, -cylSize + wallSize, 0, nearCylDensity};
37
38   Circle(1)  = {2, 1, 3};    Circle(2)  = {3, 1, 4};
39   Circle(3)  = {4, 1, 5};    Circle(4)  = {5, 1, 2};
40   Circle(5)  = {18, 1, 19};  Circle(6)  = {19, 1, 20};
41   Circle(7)  = {20, 1, 21};  Circle(8)  = {21, 1, 18};
42   Circle(9)  = {6, 1, 7};    Circle(10) = {7, 1, 8};
43   Circle(11) = {8, 1, 9};    Circle(12) = {9, 1, 6};
44   Circle(13) = {14, 1, 15};  Circle(14) = {15, 1, 16};
45   Circle(15) = {16, 1, 17};  Circle(16) = {17, 1, 14};
46   Circle(17) = {10, 1, 11};  Circle(18) = {11, 1, 12};
47   Circle(19) = {12, 1, 13};  Circle(20) = {13, 1, 10};
48
49   Line Loop(1) = {4, 1, 2, 3};
50   Plane Surface(1) = {1};
51   Line Loop(2) = {5, 6, 7, 8};
52   Plane Surface(2) = {1, 2};
53   Line Loop(3) = {9, 10, 11, 12};
54   Plane Surface(3) = {2, 3};
55   Line Loop(4) = {13, 14, 15, 16};
56   Plane Surface(4) = {3, 4};
57   Line Loop(5) = {17, 18, 19, 20};
58   Plane Surface(5) = {4, 5};
59   Physical Surface("out") = {4, 5};
60   Physical Surface("in") = {1, 2};
61   Physical Surface("wall") = {3};
```

A part of particular interest is three lines 59–61 that say `Physical Surface`. They mark three physical regions: inside the cylinder, the wall and the outside. We will use them later in FEniCS to integrate over these three regions separately. Figure 2 shows an example of a generated mesh without keeping the real proportions that are specified in the script above, the real proportions would hinder the details. Giving that the script name is `sc.geo`, you can use the following commands to generate the FEniCS files:

```
gmsh -2 sc.geo
dolfin-convert sc.msh sc.xml
```

This will generate an XML definition of the mesh `sc.xml` and an XML file `sc_physical_region.xml` that labels the cells according to the `Physical Surface` statements in the `sc.geo` file. The text labels are not supported in the Python code, they are just consequent numbers corresponding to the `Physical Surface` statements order. We just have to memorize that 1 corresponds to the outside of the cylinder, 2 to the inside, and 3 to the wall.

### FEniCS code

All code assumes the following imports:

```python
from fenics import *
from scipy.special import kn
import matplotlib.pyplot as plt
import math
```

The first step is to read the generated mesh and define a function space (line numbers in the listings are given just for a reference, they don't reflect numbers in a real program).

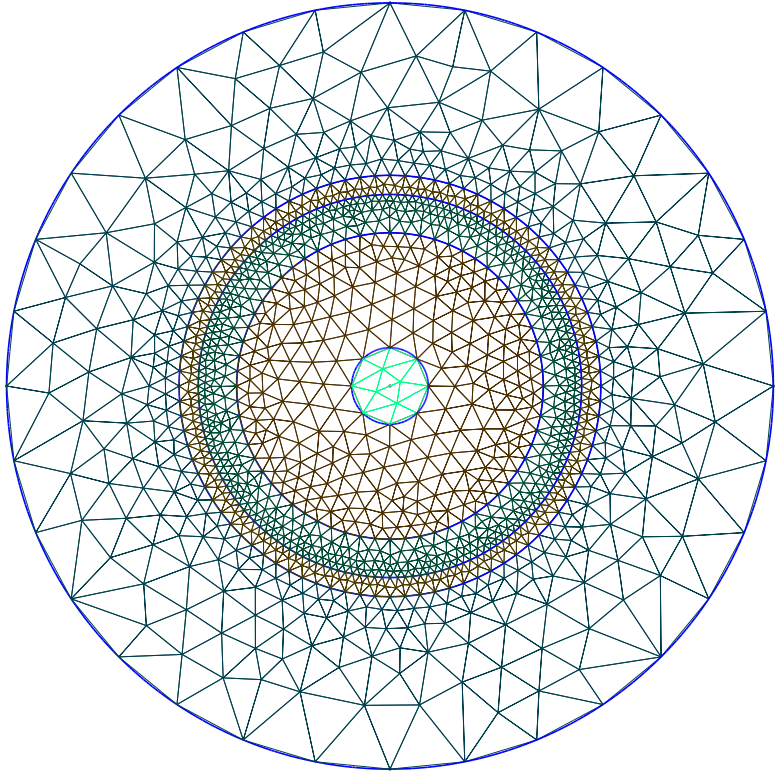

**Figure 2:** An example of mesh generated by a Gmsh script. The mesh is denser near the boundary of the cylinder that holds the metal stripes. Differently colored regions show areas of the mesh created to achieve mesh density variation. The green ring area corresponds to the cylinder wall.

```python
mesh = Mesh("sc.xml")
# sc_physical_region.xml has a number assigned to each cell of sc.xml. 1 for the
# outer region of the cylinder, 2 for the inside, and 3 for the cylinder wall
domains = MeshFunction("size_t", mesh, "sc_physical_region.xml")
outDom  = 1 # mesh function value for outside of the cylinder
inDom   = 2 # mesh function value for the inside of the cylinder
wallDom = 3 # mesh function value for the cylinder wall

# The following variables are global properties of our mesh.
cylRadius = 1
outsideRadius = 5
dx = Measure("dx", domain = mesh, subdomain_data = domains)
ε = {
      outDom  : Constant(1.0), # air
      inDom   : Constant(1.0), # air
      wallDom : Constant(10.0) # glass
    }

# Function space defines how the solution is represented within each mesh cell.
# CG means Lagrange interpolation polynomials, the last argument is the order.
V = FunctionSpace(mesh, 'CG', 2)
```

Then we write classes that will define the location of the metal plates and the outside edge of the mesh.

```python
class TopPlate(SubDomain):
    def __init__(self, cylRadius, plateWidth, ω):
        self.cylRadius = cylRadius
        self.maxX = (cylRadius *
                     math.sin(plateWidth / cylRadius / 2 *
                              math.sqrt(1 + (ω * cylRadius) ** 2)))
        SubDomain.__init__(self)
    def inside(self, x, on_boundary):
```

```
30              r = math.sqrt(x[0] * x[0] + x[1] * x[1])
31              return (near(r, self.cylRadius, 0.001) and x[1] > 0 and
32                      between(x[0], (-self.maxX, self.maxX)))
33
34     class BottomPlate(SubDomain):
35         def __init__(self, cylRadius, plateWidth, ω):
36             self.cylRadius = cylRadius
37             self.maxX = (cylRadius *
38                          math.sin(plateWidth / cylRadius / 2 *
39                                   math.sqrt(1 + (ω * cylRadius) ** 2)))
40             SubDomain.__init__(self)
41         def inside(self, x, on_boundary):
42             r = math.sqrt(x[0] * x[0] + x[1] * x[1])
43             return (near(r, self.cylRadius, 0.001) and x[1] < 0 and
44                     between(x[0], (-self.maxX, self.maxX)))
45
46     class OutsideCircle(SubDomain):
47         def __init__(self):
48             SubDomain.__init__(self)
49         def inside(self, x, on_boundary):
50             return on_boundary
```

These classes inherit from the FEniCS class `SubDomain`. We redefine one function inside, which returns true if x is "inside" a domain. Since we will use these classes to define the edges of boundaries, "inside" means close enough to an edge of the mesh. That is why we use a function near, which returns `True` if the first two arguments are near each other within the tolerance specified by the third argument. The calculation for the top and the bottom plates look complicated. Here what they are. If we take a stripe of thin foil of width $d$ and wrap it around a cylinder with radius $r$ with frequency $\omega$, then in the cross section of the cylinder our foil will occupy an angle (imagine a pie diagram with a section enclosing the metal stripe)

$$\alpha = \frac{d}{r}\sqrt{1+\omega^2 r^2}. \tag{20}$$

A sine of half of it multiplied by $r$ is the maximum $x$ coordinate by module that our stripe occupies. Therefore, our "inside" condition is $x^2 + y^2$ is near $r^2$ and $x$ is between $-r\sin\alpha/2$ and $r\sin\alpha/2$. For the outside edge we simply return `on_boundary`, which corresponds to the edge of the mesh.

Next, we can use these classes to mark the edges of the mesh according to the region they belong to:

```
51     # MeshFunction assigns numbers to the mesh elements. The last argument of
52     # the MeshFunction constructor is the dimension of the mesh elements over
53     # which this function is defined. 1 means the mesh's edges.
54     boundaries = MeshFunction("size_t", mesh, 1)
55     boundaries.set_all(0)
56     TopPlate(r, d, ω).mark(boundaries, top)
57     BottomPlate(r, d, ω).mark(boundaries, bottom)
58     OutsideCircle().mark(boundaries, outside)
```

Now the marked edges can be used to define the Dirichlet boundary conditions:

```
59     # DirichletBC is a class for the Dirichlet boundary conditions. The first
60     # argument of the constructor is the function space. The second argument is
61     # an expression for the potential values on the boundary. In our case it is
62     # a simple constant. The third argument is a mesh function that marks some
63     # edges. The last argument is the value of the mesh function on the edges
64     # that belong to the boundary.
65     topPlateDC = DirichletBC(V, Constant(0.5), boundaries, top)
66     bottomPlateDC = DirichletBC(V, Constant(-0.5), boundaries, bottom)
```

Just a reminder from the FEniCS tutorial. The way we specify $a$ and $L$ from equations (18) and (19) in Python is we right down the expressions under the integrals with special variables dx and ds that represent differential elements for the volume and surface integrals correspondingly. Both dx and ds are arrays in our case, each element corresponds to some area inside the mesh. For example, ds(3) can correspond to a surface integral over edges of the mesh that are marked with number 3. This ds can be defined as following:

```
67    # This is a differential for a surface integral. We need it to define the
68    # following integral for a Rubin boundary condition in the bilinear form.
69    # $\oint_{\partial\Gamma_R} vu_{2D} \frac{\omega(K_0(\omega R)+K_2(\omega R))}{2K_1(\omega R)} ds$
70    ds = Measure('ds', domain = mesh, subdomain_data = boundaries)
```

Next, we calculate the expression under the surface integral:

```
71    # This is an expression for the potential derivative along the radius. Since
72    # at zero $K_n$ converges to infinity, for small arguments the expression
73    # switches to a boundary condition that would be for a very small dipole.
74    # $\frac{\omega(K_0(\omega R)+K_2(\omega R))}{2K_1(\omega R)}$
75    knArg = ω * R;
76    outsideDeriv = (1.0 / R if knArg < 0.01 else
77                    0.5 * ω * (kn(0, knArg) + kn(2, knArg)) / kn(1, knArg))
```

Vector $\nabla_\omega$ in Cartesian coordinates (equation (6)) will look like this:

```
78    g = as_vector((
79            (1 + ω ** 2 * x[1] ** 2) * u.dx(0) -       ω ** 2 * x[0] * x[1] * u.dx(1),
80             -ω ** 2 * x[0] * x[1] * u.dx(0) + (1 + ω ** 2 * x[0] ** 2) * u.dx(1)
81        ))
```

Finally, we can write down the linear and the bilinear forms:

```
82    a = (dot(ε[outDom]  * g, grad(v)) * dx(outDom) +
83         dot(ε[inDom]   * g, grad(v)) * dx(inDom) +
84         dot(ε[wallDom] * g, grad(v)) * dx(wallDom) +
85         v * u * outsideDeriv * ds(outside))
86    L = f * v * dx(outDom) + f * v * dx(inDom) + f * v * dx(wallDom)
```

where $\varepsilon$ is a dielectric constant. Putting everything together here is the full code for a function that calculates the potential:

```
87    def Potential(
88            ω,       # angle of spiral rotation per a unit of length
89            ε,       # dielectric constants inside and outside of the cylinder
90            r,       # cylinder radius
91            d,       # plates width
92            R,       # radius of the whole area we are solving the Laplace equation
93                     # in
94            V,       # function space
95            mesh,    # mesh for the finite lements method
96            dx       # area measurement with marked regions for the inside and
97                     # outside of the cylinder
98        ):
99        u = TrialFunction(V)
100       v = TestFunction(V)
101       f = Constant(0.0)
102       x = SpatialCoordinate(mesh)
103
104       top     = 1 # an index for the top plate mesh edges
105       bottom  = 2 # an index for the bottome plate mesh edges
106       outside = 3 # an index  for the outside boundary mesh edges
107
108       boundaries = MeshFunction("size_t", mesh, 1)
109       boundaries.set_all(0)
110       TopPlate(r, d, ω).mark(boundaries, top)
111       BottomPlate(r, d, ω).mark(boundaries, bottom)
112       OutsideCircle().mark(boundaries, outside)
113
114       topPlateDC = DirichletBC(V, Constant(0.5), boundaries, top)
115       bottomPlateDC = DirichletBC(V, Constant(-0.5), boundaries, bottom)
116
117       ds = Measure('ds', domain = mesh, subdomain_data = boundaries)
118
119       knArg = ω * R;
```

```
120        outsideDeriv = (1.0 / R if knArg < 0.01 else
121                        0.5 * ω * (kn(0, knArg) + kn(2, knArg)) / kn(1, knArg))
122        # Redefine ω as a FEniCS expression.
123        ω = Constant(ω)
124        g = as_vector((
125              (1 + ω ** 2 * x[1] ** 2) * u.dx(0) -     ω ** 2 * x[0] * x[1] * u.dx(1),
126                  -ω ** 2 * x[0] * x[1] * u.dx(0) + (1 + ω ** 2 * x[0] ** 2) * u.dx(1)
127           ))
128        a = (dot(ε[outDom]  * g, grad(v)) * dx(outDom) +
129             dot(ε[inDom]   * g, grad(v)) * dx(inDom) +
130             dot(ε[wallDom] * g, grad(v)) * dx(wallDom) +
131             v * u * outsideDeriv * ds(outside))
132        L = f * v * dx(outDom) + f * v * dx(inDom) + f * v * dx(wallDom)
133
134        u = Function(V)
135        solve(a == L, u, [topPlateDC, bottomPlateDC])
136
137        return u
```

We are almost there. We have to find the energy density, which is defined by the equation (9). We can do it by using the same solver `solve(a == L, ...)`:

```
138    def EnergyDensity(pot, ω, ε, V, mesh, dx, outDom, inDom):
139        u = TrialFunction(V)
140        v = TestFunction(V)
141        x = SpatialCoordinate(mesh)
142
143        de = dot(grad(pot), grad(pot)) + ω ** 2 * (x[1] * pot.dx(0) + x[0] * pot.dx(1)) ** 2
144        a = u * v * dx(outDom) + u * v * dx(inDom)
145        L = (ε[outDom] * de * v * dx(outDom) +
146             ε[inDom]  * de * v * dx(inDom))
147
148        energyDensity = Function(V)
149        solve(a == L, energyDensity)
150
151        return energyDensity
```

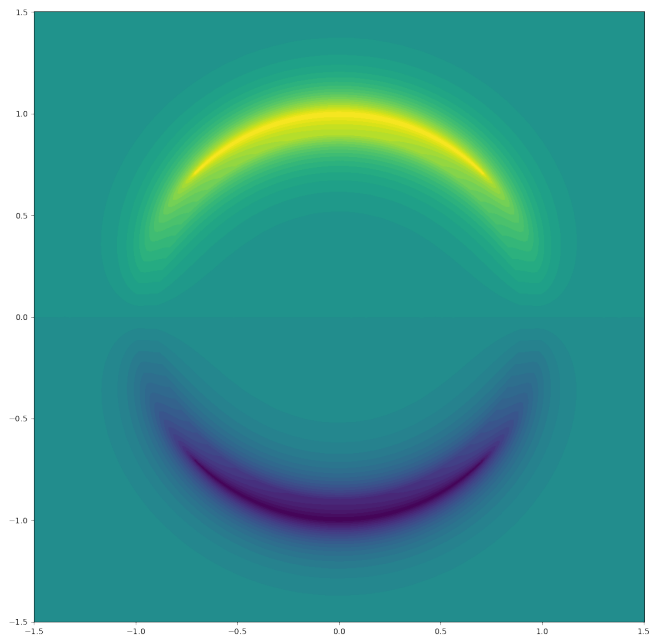

**Figure 3:** Electrostatic potential output example of the FEniCS program.

The following code will show a color map of the potential and will calculate the energy stored in the capacitor:

```
152    def main():
153        plateWidth = 0.2
154
155        u2D = lambda ω: Potential(ω, ε, cylRadius, plateWidth, outsideRadius, V,
156                                  mesh, dx)
157        E = lambda u, ω: EnergyDensity(u, ω, ε, V, mesh, dx)
158
159        max_ω = math.sqrt((math.pi / plateWidth) ** 2 - (1.0 / cylRadius) ** 2)
160
161        ω = max_ω / 2
162        u = u2D(ω)
163
164        plt.figure()
165        plot(u)
166        plt.show()
167
168        energyDensity = E(u, ω)
169        inEnergy = assemble(energyDensity * dx(inDom))
170        outEnergy = (assemble(energyDensity * dx(outDom)) +
171                     assemble(energyDensity * dx(wallDom)))
172        insidePart = inEnergy / (inEnergy + outEnergy)
173
174        print("Inside energy ({omega}) = {energy}".format(omega = ω, energy = inEnergy))
175        print("Outside energy ({omega}) = {energy}".format(omega = ω, energy = outEnergy))
176        print("Total energy ({omega}) = {energy}".format(omega = ω, energy = inEnergy + outEnergy))
177        print("Inside energy part ({omega}) = {energy}".format(omega = ω, energy = insidePart))
178
179        return 0
180
181    if __name__ == '__main__':
182        exit(main())
```

# Experiments and results

It's time to conduct some numeric experiments to answer the ultimate question: what is the best way to design the system? Before we can venture into answering this, we need to acquire a general sense of how the system behaves, i.e. what are the physical consequences of different choices.

First, let's get some impression how the potential behaves when we change $\omega$. Figure 4 shows the potential along the vertical line through the center of the cylinder cross-section for different $\omega$. Since the plate width in the cross section increases with $\omega$, the maximum $\omega$ corresponds to $\alpha$ in (20) equal to $\pi$.

$$\omega_{max} = \sqrt{\frac{\pi^2}{d^2} - \frac{1}{r^2}}$$

We see that as we increase the winding angle, the potential starts quickly decreasing father away from the metal plate. Since the energy is proportional to the squared gradient of the potential, we can expect that the capacitance will be increasing with $\omega$ as well.

Let's set the plate width to some small number and calculate the capacitance dependence on $\omega$ without modifying the plate width for each $\omega$ (by removing `math.sqrt(1 + (ω * cylRadius) ** 2)` from the `TopPlate` and `BottomPlate` classes, lines 27 and 39). I.e. this will be a dependency purely on only one parameter $\omega$ in the equation. The result presented in fig. 5. The winding frequency is in loops per length equal to radius, i.e. the $X$ axis of the chart is $\omega R/2\pi$, where $R$ is the radius of our cylinder. As expected, the capacitance grows with the winding angle. Moreover, it grows very fast, almost exponentially. Therefore, one might think that the best strategy is to take a very narrow metal stripe and wind it around the cylinder as tightly as we can. However, it's not the right answer.

We should step back and contemplate on what we are trying to build. We want our capacitance to change depending on the presence of water inside. Water has high dielectric constant 81. I.e. it will push out the electric field from the inside of the cylinder. Since the capacitance is equivalent to the stored energy, and the stored energy is the electric field squared, the effect of the water is roughly in pushing the part of the stored energy from the inside of the cylinder to the inside of its wall. This argument may sound a little counterintuitive, therefore we are going to demonstrate it with a following simple example. Let's take an unfussy plane capacitor: two flat plates, separated by a gap. Assuming the gap size is 1 in some arbitrary units and omitting multiplicative constants due to the choice of dimensions, the capacitance is $C = \epsilon S$, where $S$ is the area of the plates, if the capacitor's media has the dielectric constant $\epsilon$. Everything as expected: we know the higher the $\epsilon$ of the media, the higher the capacitance. However, let's see what happens if there is a small breathing space $\delta$ between the plates and the dielectric. The electric field is perpendicular to the interface between the dielectric and vacuum and $D_\perp$ should be continuous, which means $\epsilon E_{media} = E_{vacuum}$. We see that the electric field inside the media is $\epsilon$ times lower, in other words the electric field was pushed out from the dielectric to the vacuum if $\epsilon \gg 1$. Giving the dielectric media thickness is $1 - \delta$, the capacitance now is $C = \epsilon S/(1 + (\epsilon - 1)\delta)$. If $\epsilon \gg 1$ and $\delta$ is not infinitesimally small, it will turn into $C = S/\delta$. This is equivalent to a capacitor without a deielectric, but with distance $\delta$ between plates. The high $\epsilon$ dielectric pushed all electric field out to the small $\delta$- gap.

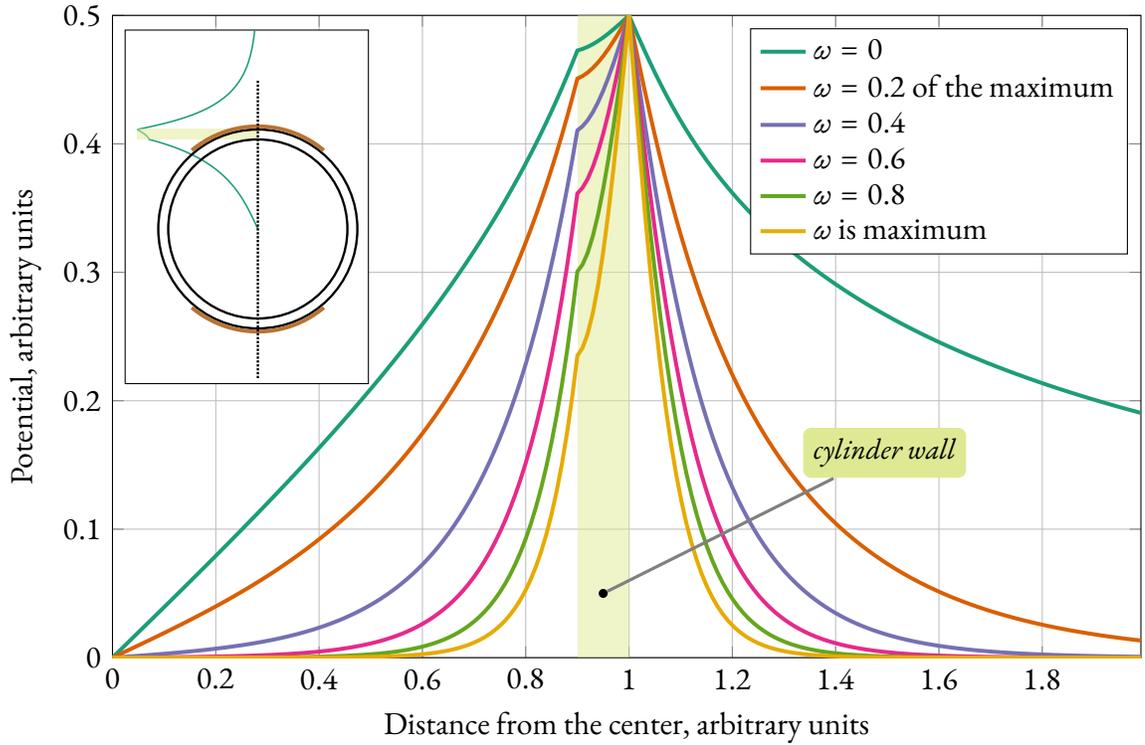

**Figure 4:** Potential along $Y$ axis (dotted line in the inlet) for different winding angles.

In our case the $\delta$-gap is the wall of the cylinder. Therefore, the sensitivity of our sensor will be the better the more percentage of the total stored energy is inside the cylinder. However, as we see from fig. 4, as we increase the winding frequency, all the field starts being concentrated in the vicinity of the plates, ignoring the inner part behind the wall. Our capacitance is for sure large, but it won't be reactive to the water inside. It's easy to understand if we imagine what is going on when we increase the winding frequency. The neighboring loops become closer and closer, along the side of the cylinder, not across, the capacitance will be mostly due to the neighboring edges, not to the contents of the cylinder.

What we have to check is not the total capacitance given by equation (11), but only the part originated from the energy stored in the inner area of the cylinder. It can be done by integrating only over the inner circle in equation (10). The program snippet above shows how to do it in line 169: `inEnergy = assem-`

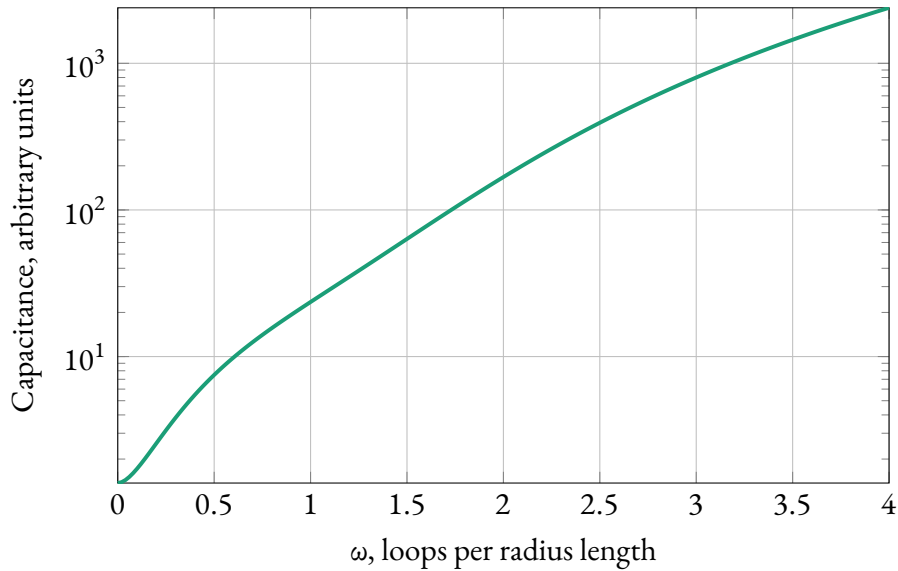

**Figure 5:** Capacitance vs winding frequency. Cylinder wall thickness $w$ = 0.025.

ble(energyDensity * dx(inDom)). Let's construct two charts over $\omega$: (i) the inner capacitance; (ii) the ratio of the inner capacitance to the total capacitance. The result is presented in fig. 6. As expected, the inner capacitance first increases, because the total capacitance grows very fast, then, as more and more energy starts being concentrated in the vicinity of the plates, the inner capacitance starts dropping. The second curve shows what part of the capacitance is concentrated in the inside of the cylinder. It roughly corresponds to the sensitivity of our sensor. It has a clear maximum around 1. This is our sweet spot, because it maximizes both the capacitance and the sensitivity. Note that in fig. 6 the wall much thinner than in fig. 4. The potential drops so rapidly inside the wall that for thicker walls the maximum in fig. 6 becomes less and less noticeable, until flattens out.

As we change the plate width, the location and the height of the sensitivity peak will change too. For example, we can expect that it will disappear completely as the plates become really wide, because then they again will interact mostly along the side of the cylinder, than across. Super thin plates will also be less effective, because they will achieve the maximum sensitivity at the larger winding angles, where the field is concentrated in too much outside, and, therefore the size of the peak will be smaller. Somewhere in the middle there should be an optimal value for both the winding frequency and the plate width. We can find it using a Python function from scipy.optimize called minimize. Here is an example of the corresponding code:

```python
def Sensetivity(x):
    insidePart = 0
    if x[1] > 0.02:
        u2D = lambda ω, plateWidth: Potential(ω / cylRadius * math.pi * 2,
                                              ε, cylRadius, plateWidth,
                                              outsideRadius, V, mesh, dx)
        E = lambda u, ω: EnergyDensity(u, ω / cylRadius * math.pi * 2, ε, V,
                                       mesh, dx)
        u = u2D(x[0], x[1])
        energyDensity = E(u, x[0])
        inEnergy = assemble(energyDensity * dx(inDom))
        outEnergy = (assemble(energyDensity * dx(outDom)) +
                     assemble(energyDensity * dx(wallDom)))
        storedEnergy = inEnergy + outEnergy
        insidePart = inEnergy / storedEnergy

    return insidePart
```

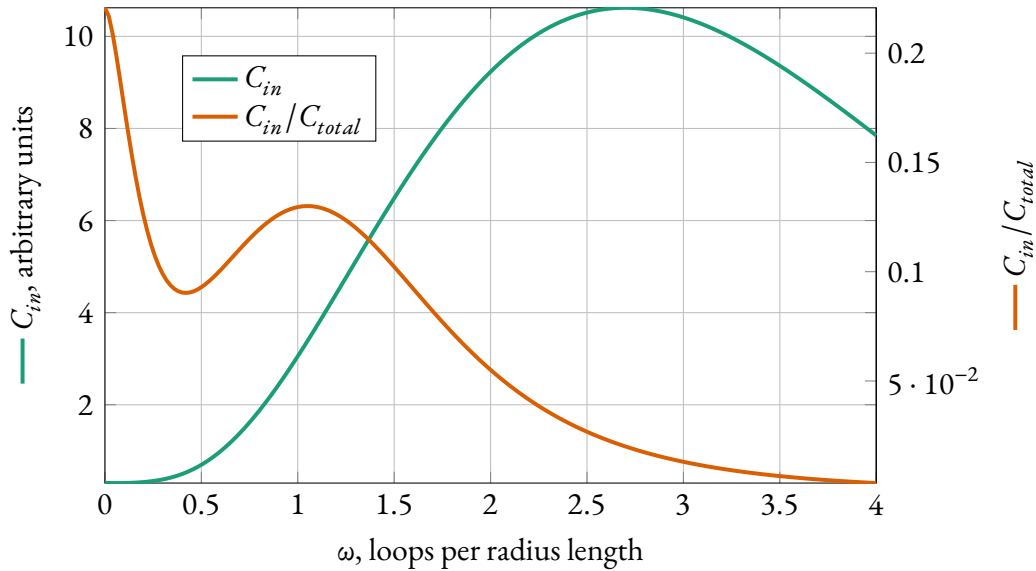

**Figure 6:** Capacitance of the inner part of the cylinder and what part of the total capacitance it is. Cylinder wall thickness $w = 0.025$.

```
201    res = minimize(lambda x : -Sensetivity(x),
202                   [0.9, 0.3335],
203                   method = 'Nelder-Mead',
204                   options = {'disp': True,
205                              'initial_simplex' : [[0.6, 0.025],
206                                                   [1.4, 0.025],
207                                                   [1, 0.6]]})
208    print(res)
```

The result of this code will be $\omega = 0.95$ and $d = 0.21$. Note that this is the plate width in cross-section of the cylinder. If we adjust it to the angle according to (20), we will get $d = 0.035$. The ratio $C_\epsilon/C_{total}$ in the maximum will be 0.15. As a final step we can calculate the sensitivity by calculating the total capacitance with and without water. We can do it by changing the dielectric constant in the ε dictionary: `ε[inDom] = 81`. For the optimum $\omega$ and $d$ the total capacitance with water will be 97, and without 37 in arbitrary units. Therefore, the capacitance will change 2.6 times for an empty and full sensor.

# Conclusion

We have a program that solves a practical task of designing a water level sensor consisting of a cylinder with spiraling stripes of metal on the sides. For the example configuration this program calculated the optimal parameters for the winding angle and the widths of the stripes.

There is one parameter we didn't vary: the thickness of the cylinder wall. Thinner the wall, more capacitance we will get. However, it is unknown how it will influence the sensitivity.

Another interesting thing is the fact that the optimal stripe width is extremely small. It's just 3.5% of the radius. In practice for a DIY project it means that it's better to glue a thin wire instead of a stripe.

Another area that needs to be treated carefully is how specify our plates in the mesh. We just mark some edges as belonging to the metal. The problem is that the electrical field in this case concentrates mostly at the edges of the metal and quickly drops several orders of magnitude. It should be investigated how it influences the precision of the calculations. We can add some shape to the planes, but that will require changing the mesh for any change in the winding angle and the width of the metal.

# References

[1]  Martin S. Alnæs et al. "The FEniCS Project Version 1.5." In: *Archive of Numerical Software* 3.100 (2015). DOI: `10.11588/ans.2015.100.20553`.

[2]  Anders Logg, Kent-Andre Mardal, Garth N. Wells, et al. *Automated Solution of Differential Equations by the Finite Element Method*. Ed. by Timothy J. Barth et al. Lecture Notes in Computational Science and Engineering 84. Springer-Verlag, 2012. ISBN: 978-3-642-23098-1. DOI: `10.1007/978-3-642-23099-8`.

[3]  FEniCS Project. *FEniCS download and installation web page*. URL: `https://fenicsproject.org/download`.

[4]  Hans Petter Langtangen and Anders Logg. *Solving PDEs in Python*. Simula SpringerBriefs on Computing 3. Springer International Publishing, 2016. ISBN: 978-3-319-52462-7. DOI: `10.1007/978-3-319-52462-7`. URL: `https://fenicsproject.org/tutorial`.

[5]  Christophe Geuzaine and Jean-François Remacle. "Gmsh: A 3-D finite element mesh generator with built-in pre- and post-processing facilities." In: *International Journal for Numerical Methods in Engineering* 79.11 (2009). DOI: `10.1002/nme.2579`.

[6]  Christophe Geuzaine and Jean-François Remacle. *Gmsh*. URL: `https://gmsh.info/doc/texinfo/gmsh.html`.